\documentclass[conference]{IEEEtran}
\usepackage{amsfonts}
\usepackage{amssymb}
\usepackage{amsmath}
\usepackage{amsthm}
\usepackage{enumerate}
\usepackage{mathrsfs}
\usepackage[caption=false]{subfig}
\usepackage[utf8]{inputenc}
\usepackage{units}
\usepackage{multirow}
\usepackage{wasysym}
\usepackage{tikz}
\usepackage{graphicx}
\usepackage{comment}
\usepackage{cite}
\usepackage[ruled,vlined]{algorithm2e}
\usepackage{cases}
\usepackage{tkz-berge}
\usepackage{todonotes}
\usepackage{url}
\newtheorem{lemma}{Lemma}
\newtheorem{definition}{Definition}

\newtheorem{theorem}{Theorem}
\newtheorem{corollary}{Corollary}

\usetikzlibrary{shadows,shapes,decorations}
\tikzstyle{pici}=[circle,draw,auto=left, minimum width=4pt, circular drop shadow, fill=white]
\tikzstyle{pont}=[circle,draw,auto=left, minimum width=15, inner sep=0pt, circular drop shadow, fill=white]
\tikzstyle{legend} = [rectangle, text width=7em, text centered]
\makeatletter
\def\pgf@plot@curveto@handler@finish{%
  \ifpgf@plot@started%
    \pgfpathcurvebetweentimecontinue{0}{0.995}{\pgf@plot@curveto@first}{\pgf@plot@curveto@first@support}{\pgf@plot@curveto@second}{\pgf@plot@curveto@second}%
  \fi%
}
\makeatother
\tikzstyle{2edgecut} =[line width=0.06cm, ->,>=stealth]
\tikzstyle{3arccut} =[->,>=stealth]
\begin{document}
\title{Resilient Flow Decomposition of Unicast Connections with Network Coding}
\author{\IEEEauthorblockN{P\'eter Babarczi\IEEEauthorrefmark{1}, J\'anos Tapolcai\IEEEauthorrefmark{1}, Lajos R\'onyai\IEEEauthorrefmark{2}, Muriel M\'edard\IEEEauthorrefmark{3}} 
\IEEEauthorblockA{\IEEEauthorrefmark{1}MTA-BME Future Internet Research Group, High-Speed Networks Laboratory (HSN\emph{Lab}),\\ 
Budapest University of Technology and Economics (BME), \{babarczi, tapolcai\}@tmit.bme.hu}
\IEEEauthorblockA{\IEEEauthorrefmark{2}Computer and Automation Research Institute Hungarian Academy of Sciences and BME, ronyai@sztaki.hu}
\IEEEauthorblockA{\IEEEauthorrefmark{3}Research Lab. of Electronics, Massachusetts Institute of Technology, Cambridge, MA 02139 USA, medard@mit.edu}
}

\maketitle
\begin{abstract}
In this paper we close the gap between end-to-end diversity coding and intra-session network coding for unicast connections resilient against single link failures. In particular, we show that coding operations are sufficient to perform at the source and receiver if the user data can be split into at most two parts over the filed $GF(2)$. Our proof is purely combinatorial and based on standard graph and network flow techniques. It is a linear time construction that defines the route of subflows $A$, $B$ and $A \oplus B$ between the source and destination nodes. The proposed resilient flow decomposition method generalizes the $1+1$ protection and the end-to-end diversity coding approaches while keeping both of their benefits. It provides a simple yet resource efficient protection method feasible in 2-connected backbone topologies. Since the core switches do not need to be modified, this result can bring benefits to current transport networks.
\end{abstract}
\begin{IEEEkeywords}
network coding, instantaneous recovery, unicast connections, resilient flow decomposition
\end{IEEEkeywords}
\section{Introduction}
\label{sec:intro}
Among the several benefits offered by in-network modification of data -- i.e., \emph{network coding} (NC)~\cite{ahlswede2000network} --, resource efficiency~\cite{babarczi2013comnet} and robustness against link failures~\cite{koetter2003algebraic,jaggi2005polynomial} are two which are important for unicast connections. In particular, they can be used for failure-protection of unicast connections in wireline transport networks by providing \emph{instantaneous recovery}, i.e., after an edge fails, the destination node is able to recover the data sent without any real-time signaling mechanism, because there is no need for flow rerouting or packet retransmission. 

The importance of instantaneous recovery has been demonstrated in several studies and in several layers of the protocol stack~\cite{babarczi2013comnet}. If only routing is considered and (network) coding is not allowed, \emph{$1+1$ protection} -- the same data flow $A$ is sent along two edge-disjoint paths -- provides sufficient resilience for the connection, while instantaneous recovery is provided with a single-ended switching at the destination node. Currently, the most resource efficient among resilience mechanisms providing instantaneous recovery is \emph{diversity coding} (DC)~\cite{babarczi2013comnet}. In DC, the source router splits the data into two parts $A$ and $B$, and sends data flows $A$, $B$ and $A \oplus B$ on three edge-disjoint paths, where $\oplus$ denotes the exclusive OR (XOR) operation. Thus, the connection is resilient against single link failures -- which are the most typical failure events in transport networks~\cite{markopoulou2004characterization} --, and the sent data can be recovered at the destination out of arbitrary two of the three data flows. However, DC is applicable only in 3-edge-connected (in the rest of the paper we refer to it simply as 3-connected) networks.

Network coding has been considered as a viable solution providing instantaneous recovery in transport networks~\cite{babarczi2013comnet,babarczi2012icc}. However, robust network codes -- where no change in the coding behavior is required after a failure occurs -- may require field size $O(|E|)$~\cite{koetter2003algebraic,jaggi2005polynomial}, which is hard to provide in practice. In order to avoid the excessive field size of the previous methods, robust network coding for unicast connections against single link failures above $GF(2)$ was proposed in~\cite{rouayheb2011robust}. This code construction method provides instantaneous recovery for shared backup protection of two unicast connections, or the backup protection of a single connection which data can be split into two parts. The main result in~\cite{rouayheb2011robust} is that the provided coding scheme for this special case is simple (i.e., only XOR coding required) compared to its previous counterparts. Furthermore, in networks which are at least 2-connected from $s$ to $t$ having some \emph{redundant} edges, but not necessarily 3-connected, it still allows instantaneous recovery in the presence of a single link failure.

In this paper, similarly to~\cite{rouayheb2011robust}, we consider the most practical scenario where the data stream is split into \emph{two parts} with equal size, respectively denoted as $A$ and $B$. Note that in practice dividing the data stream into more than two does not reduce further the required bandwidth resources due to the low connectivity at the physical layer topology \cite{babarczi2013drcn}.  Our \emph{resilient flow decomposition (RFD)} method can decompose the data stream into three end-to-end sub-flows, which are directed acyclic graphs ($A$, $B$ and $A \oplus B$), from which at least two connects the source and destination node even when a single link failure occurs. With our main result (Theorem~\ref{mainthm}), we make several steps towards the practical implementation of a resilient network coding based protection approach in transport networks. First, we demonstrate that, for two data parts, a simple strategy that doesn't require the modifications of the intermediate nodes of the network in combination with simple end-to-end coding over $GF(2)$ suffices to reap the benefits of network coding. Second, this end-to-end coding approach generalizes $1+1$ protection in networks with scarce bandwidth resources and the resource efficient diversity coding approach for 2-connected networks. Note that most backbone networks fall within this category. Finally, we give a new structure theorem (Theorem~\ref{mainthm}) for feasible networks which works linear time in the input size, in particular it works in time $O(|V|)$ for sparse networks, unlike~\cite{rouayheb2011robust} which uses a complicated network transformation, which requires $O(|V|^2)$ time complexity. 

The rest of the paper is organized as follows. Section~\ref{sec:model} introduces our problem formulation and the preliminaries. As the main finding of our paper, Section~\ref{sec:proof} provides the proof that the arcs of a feasible solution can be decomposed into three disjoint arc sets, while the corresponding algorithm is presented in Section~\ref{sec:noncritical}. Finally, Section~\ref{sec:conclusions} concludes the paper.
\section{Model and Preliminaries}
\label{sec:model}
\subsection{Problem Formulation}
\begin{table}
\centering
\caption{\label{tab:notations}Notation list for the Resilient Flow Decomposition Problem}
\begin{tabular}{|c|c|}
\hline
Notations& Description \\
\hline
\hline
$G = (V,E,k)$ & directed graph with nodes $V$, edge set $E$,  \\
 &  and free capacity $k(e) \in \mathbb{N}$ \\
\hline
$G = (V,E,c)$ & directed coding graph with nodes $V$, edge set $E$,  \\
 &  and critical capacity $c(e)$, where $c:E\rightarrow \{1,2\}$ \\
 \hline
$G^* = (V,E^*)$ & directed coding graph with nodes $V$, arc set $E^*$, where \\ 
 & each edge in $E$ is replaced by $c(e)$ parallel arcs \\
 \hline
$G' = (V,E,\bar{c})$ & reduced capacity graph with nodes $V$, edge set $E$, \\
 &    and reduced capacity $\overline{c}(e)$ \\
\hline
	$s,t$   & source and target node of the connection request \\
\hline
	$A,B$   & the two parts in which the user data is decomposed  \\ 
\hline
\end{tabular}
\end{table}

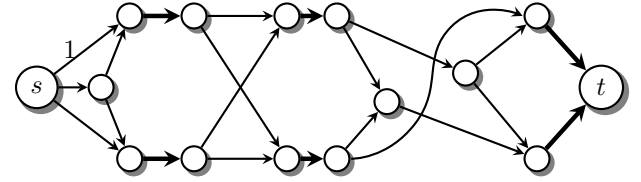
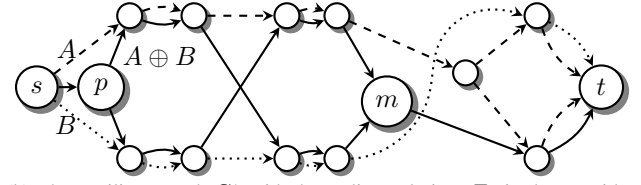
\begin{figure}[t]
\centering
\subfloat[The input graph $G$, and the reduced capacity graph $G'$. All the thin edges have $1$ unit of capacity, while the thick edges have 2 in the critical graph $G$, and $1.5$ in the reduced capacity graph $G'$.]{
  \begin{tikzpicture}[thick,scale=0.095]
      \node[pici] (s) at (3,10){$s$};
      \node[pici] (a0) at (16,20){};
      \draw[3arccut] (s) -- (a0) node [midway, left]{$1$};
       \node[pici] (b0) at (12,10){};
      \draw[3arccut] (s) -- (b0);
      \node[pici] (c0) at (16,0){}; 
      \draw[3arccut] (s) --(c0);
	\draw[3arccut] (b0) -- (c0);
	\draw[3arccut] (b0) -- (a0);
      \node[pici] (a1) at (25,20){}; 
	\draw[2edgecut] (a0) -- (a1);
      \node[pici] (c1) at (25,0){}; 
	\draw[2edgecut] (c0) --  (c1){};
      \node[pici] (a2) at (38,20){}; 
      \node[pici](c2) at (38,0){}; 
	\draw[3arccut] (a1) -- (a2);
	\draw[3arccut] (a1) -- (c2);
       \draw[3arccut] (c1) -- (c2);
       \draw[3arccut] (c1) -- (a2);
       \node[pici] (a3) at (45,20){} ;
 \draw[2edgecut] (a2) -- (a3);
       \node[pici] (c3) at (45,0){};
 	\draw[2edgecut] (c2) -- (c3);
       \node[pici] (b3) at (52,8){}; 
 	\draw[3arccut] (a3) --(b3);
 	\draw[3arccut] (c3) -- (b3);
       \node[pici] (a4) at (63,12){};
       \draw[3arccut] (a3) -- (a4);
       \node[pici] (b4) at (73,0){}; 
	\draw[3arccut] (b3) -- (b4);
	\draw[3arccut] (a4) -- (b4);
       \node[pici] (d4) at (73,20){};
       \draw[3arccut] (a4) -- (d4);
 	\draw[->,>=stealth]  plot [smooth,tension=1]  coordinates {(47,0)  (56,5)  (61,18.5)  (71,20)};
 \node[pici](d) at (82,10){$t$};
 \draw[2edgecut] (d4) -- (d);
 \draw[2edgecut] (b4) -- (d);
\end{tikzpicture}}
	\\
\subfloat[The auxiliary graph $G^*$ with the coding solution. $E_1$ is shown with broken, $E_2$ with dotted, and $E_3$ with solid lines.]{
\centering
    \begin{tikzpicture}[thick,scale=0.095]
      \node[pici] (s) at (3,10){$s$};
      \node[pici] (a0) at (16,20){};
      \draw[3arccut, dashed] (s) -- (a0) node [midway, left]{$A$};
       \node[pici] (b0) at (12,10){$p$};
      \draw[3arccut] (s) -- (b0);
      \node[pici] (c0) at (16,0){}; 
      \draw[3arccut, dotted] (s) --(c0) node [midway, left]{$B$};
	\draw[3arccut] (b0) -- (c0);
	\draw[3arccut] (b0) -- (a0) node [pos=0.2, right]{$A \oplus B$};
      \node[pici] (a1) at (25,20){}; 
	\draw[3arccut, dashed]  (a0) to [bend left=20]  (a1);
	\draw[3arccut]  (a0) to [bend right=20]  (a1);
	\node[pici] (c1) at (25,0){}; 
	\draw[3arccut]  (c0) to [bend left=20]  (c1);
	\draw[3arccut, dotted]  (c0) to [bend right=20]  (c1);
      \node[pici] (a2) at (38,20){}; 
      \node[pici](c2) at (38,0){}; 
	\draw[3arccut,dashed] (a1) -- (a2);
	\draw[3arccut] (a1) -- (c2);
       \draw[3arccut,dotted] (c1) -- (c2);
       \draw[3arccut] (c1) -- (a2);
       \node[pici] (a3) at (45,20){} ;
	\draw[3arccut, dashed]  (a2) to [bend left=20]  (a3);
	\draw[3arccut]  (a2) to [bend right=20]  (a3);     
	\node[pici] (c3) at (45,0){};
	\draw[3arccut]  (c2) to [bend left=20]  (c3);
	\draw[3arccut, dotted]  (c2) to [bend right=20]  (c3);    
	\node[pici] (b3) at (52,8){$m$}; 
 	\draw[3arccut] (a3) --(b3);
 	\draw[3arccut] (c3) -- (b3);
       \node[pici] (a4) at (63,12){};
       \draw[3arccut,dashed] (a3) -- (a4);
       \node[pici] (b4) at (73,0){}; 
	\draw[3arccut] (b3) -- (b4);
	\draw[3arccut,dashed] (a4) -- (b4);
       \node[pici] (d4) at (73,20){};
       \draw[3arccut,dashed] (a4) -- (d4);
 	\draw[->,>=stealth,dotted]  plot [smooth,tension=1]  coordinates {(47,0)  (56,5)  (61,18.5)  (71,20)};
 \node[pici](d) at (82,10){$t$};
	\draw[3arccut,dotted]  (d4) to [bend left=20]  (d);
	\draw[3arccut,dashed]  (d4) to [bend right=20]  (d);    
 	\draw[3arccut, dashed]  (b4) to [bend left=20]  (d);
	\draw[3arccut]  (b4) to [bend right=20]  (d);   
\end{tikzpicture}}
  \caption{The input topology and auxiliary graphs for the RFD problem.}\label{fig:nc2input}
\end{figure}

Let $G=(V,E,k)$ be a directed graph representing the input network of the routing problem. $V$ is the set of nodes $E$ is the set of edges (also called links), and $k(e) \in \mathbb{N}$ are the available free capacities along the edges. The network $G$ has two distinguished nodes $s$ and $t$, the source and the destination node of the unicast connection, respectively. We assume that the connection can be routed as two parts of equal size, denoted by $A$ and $B$ (for the sake of simplicity, the problem can be scaled to have both with rate 1). We assume further, that coding graph $G$ satisfies the {\em feasibility condition}: there is an $s-t$ flow of value at least 2, even if we delete an edge of $G$. We consider also critical networks for the preceding property. A feasible coding network $G=(V,E,c)$ is {\em critical}, if after deleting an edge, or decreasing the critical capacity $c(e)$ of an edge, the resulting network is no longer feasible. See also Fig.~\ref{fig:nc2input} for an illustration of a critical network. Note that if $G$ is critical, then the capacities are $c:E\rightarrow \{1,2\}$. The notation is summarized in Table~\ref{tab:notations}.

\emph{The goal in RFD is to provide instantaneous recovery in a critical $G=(V,E,c)$, i.e., to survive any single edge failure with minimal bandwidth requirement when user data can be split into two parts}, using network coding. In Section~\ref{sec:proof} we show that coding should be performed only in the source and destination nodes. It means we can route three data flows $A$ (along edges $E_1$), $B$ (along $E_2$) and $A\oplus B$ (along $E_3$) from $s$ to $t$. Our results suggest, that we only need two additional functionalities at the intermediate nodes over routing. The first one is a \emph{$1-$to$-2$ splitter} $p$, which duplicates the packets arriving on its incoming interface, and send the copies on different outgoing interfaces to add redundancy to the flow. The second one is a \emph{$2-$to$-1$ merger} $m$, which is able to switch (in a failureless state) between two identical copies of data and forward only one of them on its outgoing link. After a failure occurs, the merger forwards the intact signal on its outgoing edges. Fig.~\ref{fig:nc2input} shows (a) a problem instance of our RFD problem and (b) its solution. 

\subsection{Definitions and Preliminaries}
A cut $C \subseteq V$ is a set of nodes that contains $s$ but does not contain $t$. The edges (arcs) of a cut $C$ are the edges (arcs) from $C$ to $V \setminus C$.

Besides $G=(V,E,c)$, in our framework we shall also use the auxiliary graph $G^*=(V,E^*)$. The node set of $G^*$ is the same as the node set of $G$, and each $e\in E$ is replaced by  $c(e)$ (parallel) arcs which have the same tail and head node as $e$.

To obtain structural results, we follow the idea from~\cite{rouayheb2011robust} of using \emph{reduced capacity function} $\bar{c}$ on $E$. The reduced capacity $\bar{c}(e)$ is $1.5$ if $c(e)=2$, and it is 1 if $c(e)=1$. We denote by $G'$ the network $(V,E,\bar{c})$. We have the following:

\begin{theorem}\label{theorem1}
\cite[Theorem 2]{rouayheb2011robust}
$G=(V,E,c)$ is feasible if and only if $G'$ has an $s-t$ flow of value 
at least 3. 
\end{theorem}
Theorem~\ref{theorem1} gives an elegant characterization of feasible networks, which is very useful both for theoretical and algorithmic purposes.

\begin{lemma}\label{lemma:critical}
If $G$ is critical, then a maximum $s-t$ flow in $G'$ has value exactly 3. 
\end{lemma}
\begin{IEEEproof} 
$G$ is feasible, hence by Theorem~\ref{theorem1} $G'$ has a flow of value at least 3. It suffices to show that it cannot have a flow with higher value. Indeed, otherwise by the Ford-Fulkerson theorem the capacity of minimal $s-t$ cuts in $G'$ would be at least $3.5$. Consider one such cut, and let $e$ be an edge of the cut with $\bar{c}(e)=1.5$. We can then decrease the capacity of $e$ to $c(e)=\bar{c}(e)=1$ and still have a feasible network  by Theorem~\ref{theorem1}. If minimal cuts do not contain edges with reduced capacity $1.5$ then every cut has capacity at least 4 in $G'$, hence an arbitrary edge can be deleted while still retaining feasibility. 
\end{IEEEproof}

\begin{lemma} \label{maxflow}
\cite[Lemma 3]{rouayheb2011robust}
Suppose that $G$ is critical. Then $G'$ has a maximum flow whose value on
the edges of $G'$ is from $\{0.5,~1,~1.5\}$. In particular no edge can have flow value 0.
\end{lemma}

\begin{definition} A cut $C$ in $G$ is called
\begin{description}
\item[2-edge-cut] \qquad if $C$ has two edges both with capacity 2.
\item[3-arc-cut] \qquad if $C$ has three edges all with capacity 1. 
\end{description}
\end{definition}

\begin{corollary}
\label{23mincuts}
Suppose that $G$ is critical, and $C$ is a minimal cut in $G'$. Then cut $C$ is either a 2-edge-cut or a 3-arc-cut in $G$. 
\end{corollary}

\begin{lemma}
\label{saturated}
Suppose that $G$ is critical, and $e$ is an edge with $c(e)=2$. Then $e$ is part of a 2-edge-cut.
\end{lemma}
\begin{IEEEproof}
It suffices to show that $e$ is a part of a minimal cut in $G'$. This holds because otherwise we could decrease $c(e)$ to 1 while  still having  feasibility.
\end{IEEEproof}

\begin{lemma} \label{dag}
\cite[Proposition 4]{rouayheb2011robust} Suppose that $G$ is critical. Then $G$ is a directed acyclic graph (DAG).  
\end{lemma}
\section{Problem Analysis}
\label{sec:proof}
\subsection{Resilient Flow Decomposition Theorem}
Building on the definitions and results of Section~\ref{sec:model} we prove the existence of 3 subgraphs in $G^*$ for a critical $G$ which shows feasibility in a strong and transparent manner. 

\begin{theorem}\label{mainthm}
Suppose that $G$ is critical. Then there are disjoint arc sets $E_1,E_2,E_3$ of $G^*$ such that for an arbitrary edge $e\in E$, after removing the corresponding arc(s) at least two of the $E_i$ connects  $s$ to $t$. 
\end{theorem}

\begin{IEEEproof}
Let $C_1,C_2,\dots, C_k$ denote a maximal chain
$$C_1 \subset C_2 \subset \dots \subset C_k$$ 
of minimum $s-t$ cuts in $G'$. 
See Fig. \ref{fig:non-crossing-cuts} as an example.

\begin{figure}
\centering
\subfloat[A set of non-crossing and non-nested minimum $s-t$ cuts ($k=5$).]{
\begin{tikzpicture}[thick,scale=0.095]
      \node[pici] (s) at (3,10){$s$};
      \node[pici] (a0) at (16,20){};
      \draw[3arccut] (s) -- (a0) node [midway, left]{$1$};
       \node[pici] (b0) at (12,10){};
      \draw[3arccut] (s) -- (b0);
      \node[pici] (c0) at (16,0){}; 
      \draw[3arccut] (s) --(c0);
	\draw[3arccut] (b0) -- (c0);
	\draw[3arccut] (b0) -- (a0);
      \node[pici] (a1) at (25,20){}; 
	\draw[2edgecut] (a0) -- (a1);
	\node[legend](la) at (20,23){$1\frac{1}{2}$};
      \node[pici] (c1) at (25,0){}; 
	\draw[2edgecut] (c0) --  (c1){};
      \node[pici] (a2) at (38,20){}; 
      \node[pici](c2) at (38,0){}; 
	\draw[3arccut] (a1) -- (a2);
	\draw[3arccut] (a1) -- (c2);
       \draw[3arccut] (c1) -- (c2);
       \draw[3arccut] (c1) -- (a2);
       \node[pici] (a3) at (45,20){} ;
 \draw[2edgecut] (a2) -- (a3);
       \node[pici] (c3) at (45,0){};
 	\draw[2edgecut] (c2) -- (c3);
       \node[pici] (b3) at (52,8){}; 
 	\draw[3arccut] (a3) --(b3);
 	\draw[3arccut] (c3) -- (b3);
       \node[pici] (a4) at (63,12){};
       \draw[3arccut] (a3) -- (a4);
       \node[pici] (b4) at (73,0){}; 
	\draw[3arccut] (b3) -- (b4);
	\draw[3arccut] (a4) -- (b4);
       \node[pici] (d4) at (73,20){};
       \draw[3arccut] (a4) -- (d4);
 	\draw[->,>=stealth]  plot [smooth,tension=1]  coordinates {(47,0)  (56,5)  (61,18.5)  (71,20)};
 	\draw[dotted]  plot [smooth cycle] coordinates {(-1,10) (2.5,13.5) (7.5,13) (8,2.5)  (2,5)};
 	\draw[dotted]  plot [smooth cycle] coordinates {(-2,11) (17,23) (19,9.5) (17,-3)  (4.5,0.5)};
 	\draw[dotted]  plot [smooth,tension=1]  coordinates {(17,23) (26,23.5)  (40,22.5)  (42,10)  (39,-2.5) (26,-3)};
 	\draw[dotted]  plot [smooth,tension=1]  coordinates {(37,24) (53.5,21) (58.5,10) (54,-2) (43.5,-3.5)};
 	\draw[dotted]  plot [smooth,tension=1]  coordinates {(51,22.5) (76,22.5) (77,9.5) (77.5,-2.5) (70,-3.5)};
 	\node[legend] (la) at (6,5) {$C_1$};
 	\node[legend] (la) at (11,-1) {$C_2$};
 	\node[legend] (la) at (23.5,-3.5) {$C_3$};
 	\node[legend] (la) at (41,-3.5) {$C_4$};
 	\node[legend] (la) at (68,-3.5) {$C_5$};
 \node[pici](d) at (82,10){$t$};
 \draw[2edgecut] (d4) -- (d);
 \draw[2edgecut] (b4) -- (d);
\end{tikzpicture}}\\
\subfloat[The corresponding subgraphs $G^*_i$ for $i=1,\dots, k-1$.]{
    \begin{tikzpicture}[thick,scale=0.095]
	\node[legend] at (5,3){$G^*_1$};
 	\node[pici](s) at (3,10){$s_1$};
       \node[pici](a0) at (16,20){};
       \draw[3arccut] (s) -- (a0) node [midway, left]{$1$};
       \node[pici](b0) at (12,10){}; 
       \draw[3arccut] (s) -- (b0);
       \node[pici](c0) at (16,0){};
       \draw[3arccut] (s) -- (c0){};
 	\draw[3arccut] (b0) -- (c0){};
 	\draw[3arccut] (b0) -- (a0){};
       \node[pici](t1) at (25,10){$t_1$}; 
 	\draw[3arccut] (a0) to [bend left=20] (t1){};
 	\draw[3arccut] (a0) to [bend right=20]  (t1){};
 	\draw[3arccut] (c0) to  [bend left=20] (t1){};	
 	\draw[3arccut] (c0) to [bend right=20] (t1){};
 \node[legend](A) at (16,-15){$G^*_2$};
       \node[pici](s2) at (16,-8){$s_2$} ;
       \node[pici](a1) at (25,2){};
 \draw[3arccut](s2) to [bend left=20]  (a1){};
 \draw[3arccut] (s2)  to [bend right=20]  (a1){};
       \node[pici](c1) at (25,-18){};
 \draw[3arccut](s2)  to [bend right=20]  (c1){};
 \draw[3arccut](s2)  to [bend left=20]  (c1){};
       \node[pici](a2) at (38,2){} ;
       \node[pici](c2) at (38,-18){}; 
 \draw[3arccut](a1) -- (a2){};
	\draw[3arccut] (a1) --(c2){};
 \draw[3arccut](c1) --(c2){};
\draw[3arccut] (c1) -- (a2);
       \node[pici](t2) at (45,-8){$t_2$}; 
 \draw[3arccut] (a2) to [bend left=20] (t2){};
 \draw[3arccut] (a2) to [bend right=20] (t2){};
 \draw[3arccut] (c2) to [bend left=20] (t2){};
 \draw[3arccut] (c2) to [bend right=20] (t2){};
   \node[legend](A) at (38,16){$G^*_3$};
 \node[pici](s3) at (43,10){$s_3$}; 
 \node[pici](a3) at (50,20){};
   \draw[3arccut,dashed] (s3) to [bend left=20] (a3);
\node[legend] at (43,19) {$f$};
\draw[3arccut] (s3) to [bend right=20] (a3);
 \node[pici](c3) at (50,0){$v$} ;
   \draw[3arccut] (s3) to [bend left=20] (c3);
\node[legend] at (43,3) {$g$};
\draw[3arccut,dotted] (s3) to [bend right=20] (c3);
 \node[pici](b3) at (58,10){$m$} ;
   \draw[3arccut](a3) -- (b3);
   \draw[3arccut](c3)  -- (b3);
   \node[legend] at (51.5,6.5) {$h_1$};
 \node[pici](t3) at (71,10){$t_3$};
   \draw[3arccut,dashed](a3) -- (t3);
   \node[legend] at (59,18) {$e_1$};
   \draw[3arccut](b3) -- (t3);
   \node[legend] at (63.5,11.5) {$e_2$};
   \draw[3arccut,dotted] (c3) -- (t3);
   \node[legend] at (60,3) {$e_3$};
\node[legend](A) at (55,-15){$G^*_4$};
 \node[pici](s4) at (55,-8){$s_4$} ;
  \node[pici](a4) at (65,-8){};
\draw[3arccut](s4) -- (a4);
       \node[pici](b4) at (73,-17){} ;
 \draw[3arccut](s4) -- (b4);
 \draw[3arccut](a4) -- (b4);
       \node[pici](d4) at (73,3){};
	\draw[3arccut](a4) -- (d4){};
 \draw[3arccut](s4) -- (d4){};
 \node[pici](d) at (82,-7){$t_4$};
   \draw[3arccut] (d4) to [bend left=20] (d);
\draw[3arccut] (d4) to [bend right=20] (d);
   \draw[3arccut] (b4) to [bend left=20] (d);
\draw[3arccut] (b4) to [bend right=20] (d);
\end{tikzpicture}}
  \caption{Illustrative example for the proof of Theorem \ref{mainthm}}\label{fig:non-crossing-cuts}
\end{figure}
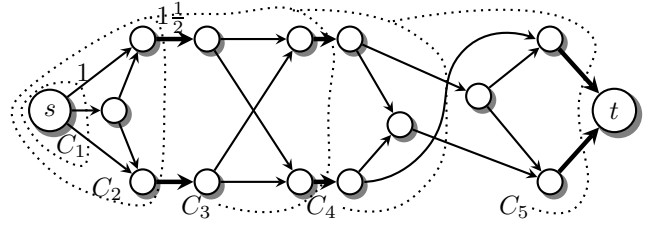
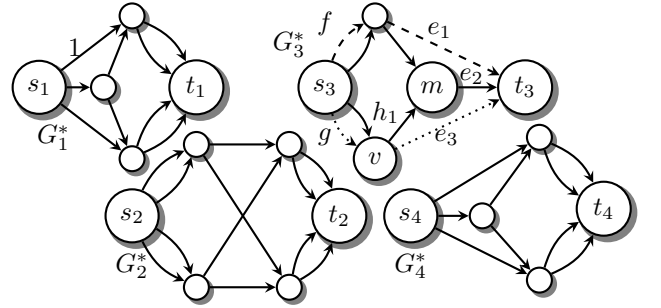

For easier understanding, we will use a set of
auxiliary graphs, denoted by $G^*_1,\dots,G^*_{k-1}$. Each $G^*_i$  has a
source node $s_i$ and a destination node $t_i$, and also 
the nodes of $C_{i+1}\setminus C_{i}$, for $i=1,\dots,k-1$.
The arcs of $G^*_i$ are the arcs between the 
nodes $C_{i+1}\setminus C_{i}$ and the arcs crossing $C_i$ but with their 
start node replaced by a new node $s_i$. Also, the arcs crossing   
$C_{i+1}$ will be added to $G^*_i$ but with their endpoint replaced by 
a new node $t_i$. See
also Fig. \ref{fig:non-crossing-cuts}(b).

Next, for each graph $G^*_i$ we define 3 disjoint arc 
sets $E^i_1,E^i_2,E^i_3$, such that every $E^i_1,E^i_2,E^i_3$
carries 1 unit of flow from $s_{i}$ to $t_{i}$.
We next show that the arc sets $E^i_1,E^i_2,E^i_3$ are 
\emph{feasible} in the sense 
that after removing the arc(s) of $G^*_i$ corresponding to an arbitrary
edge $e\in E$, at least two of the $E^i_1,E^i_2,E^i_3$ still connects
$s_{i}$ to $t_{i}$.

Finally we indicate how to piece together $E_1,E_2,E_3$ from the 
local pieces $E^i_j$.

Our argument takes $G^*_1,G^*_2,\dots,G^*_{k-1}$ one by 
one, and proves the existence of a feasible solution for each.
According to Corollary \ref{23mincuts}, $C_i$ is either a 3-arc-cut 
or a 2-edge-cut, therefore for $G^*_i$ we have four cases to consider:
\begin{description}\setlength{\labelwidth}{4em}\setlength{\itemindent}{2em}
\item[Type (i)] Both $C_i$ and $C_{i+1}$ are 3-arc-cuts.
\item[Type (ii)] Both $C_i$ and $C_{i+1}$ are  2-edge-cuts.
\item[Type (iii)] $C_i$ is 2-edge-cut and $C_{i+1}$ is 3-arc-cut.
\item[Type (iv)] $C_i$ is 3-arc-cut  and $C_{i+1}$ is 2-edge-cut.
\end{description}


We note beforehand that every graph $G^*_i$ inherits a flow of value 3 from 
$G'$, and they have only two minimum cuts with respect to the reduced
capacities, namely $\{s_i\}$ and $\{s_i\}\cup C_{i+1}\setminus C_i$. Also they are 
critical graphs (otherwise $G$ would not be critical). These with Lemma 
\ref{saturated} 
imply that 
except for the arcs incident to $s_i$ and $t_i$ they can not have arcs corresponding to edges
with reduced capacity $1.5$. 

We remark also that the three arcs of a 3-arc-cut $C_i$ must be 
one-to-one correspondence with $E^i_1,E^i_2,E^i_3$, whereas for a 2-edge-cut 
we must have a \emph{dominating} arc set, which has arcs from both edges.
This dominant set will always be denoted by $E^i_1$. 

We now consider Types~(i)-(iv) one by one. The only complicated cases are Type~(iii) and Type~(iv). In Type (i), 
$G^*_i$ can not have arcs corresponding to reduced capacity $1.5$ hence 
the reduced capacities in $G^*_i$ are integral. This implies that there are 
three arc disjoint paths $P_1,P_2,P_3$ from
$s_i$ to $t_i$ in $G^*_i$ and these 3 paths can be set as $E^i_1,
E^i_2,E^i_3$. 

For Type~(ii) there must be four arc disjoint paths $P_1,P_2,P_3, P_4$
between $s_{i}$ to $t_{i}$, because with the original capacities 
the value of minimal cuts in $G^*_i$ is 4, and the edges in $C_{i+1}\setminus C_i$ all 
have  capacity 1 in this setting by Lemma~\ref{saturated}. We intend to form $E^i_1,E^i_2,E^i_3$  by 
taking
the union of  a suitably chosen path pair $P_a,P_b$,  ($a\not=b$) as $E^i_1$, and
$E^i_2,E^i_3$ will be the remaining two paths. We have to make sure that for
the selected pair $(a,b)$ $P_a$ and $P_b$ do not contain the two arcs of the
same edge of capacity 2, because in this case $E^i_2,E^i_3$ has to traverse the 
other edge in the 2-edge-cut, and the failure of the latter edge would disconnect both $E^i_2,E^i_3$. There are at most 4 such edges, the edges of $C_i$ 
and $C_{i+1}$, and one such edge rules out at most one pair $(a,b)$. Thus,  
among the 6 possible path pairs there will be a suitable one. 

For Type~(iii) the (capacity 2) edges $f,g$ are adjacent to $s_{i}$, and 
$e_1,e_2, e_3$ are the arcs adjacent to $t_{i}$. See also $G^*_3$ on 
Fig.~\ref{fig:non-crossing-cuts}(b) for illustration. 
We will show there 
are three arc disjoint subgraphs, such that two of them are 
paths traversing $f \rightarrow e_x$, and $g \rightarrow e_y$ 
where $x\neq y$, while the third subgraph is composed of 
three path segments $f \rightarrow m$, $g \rightarrow m$, and $m \rightarrow e_z$, 
where $m$ is called merger node and $z\neq y$ and $z\neq x$.
To prove it we need to dig into Ford-Fulkerson
theorem.
Let $F$ be an integer    
flow of value 3 in $G^*_i$. The existence of this flow implies that there are 
three arc disjoint 
paths $P_1,P_2,P_3$ from $s_i$ to $t_i$ in $G^*_i$. The arcs
$e_1,e_2, e_3$, and 3 of the 4 arcs corresponding to $f,g$ are traversed by 
$F$ because they originate from minimum 
cuts in $G'$. Without loss of generality we may assume that $P_1$ and $P_2$ 
traverses the arcs in $G^*_i$ corresponding to $f$, and $P_3$ traverses $g$    
and hence passes through node $v$, where $g=(s_i,v)$. There must exist 
a $v-t_i$ flow in $G^*_i$ of value at least 2, for otherwise $G^*_i$ would    
not be feasible (remove $f$).

Let $P$ be the $v-t_i$ path $P_3\setminus g$ (e.g. it is $e_3$ on $G^*_3$).
We can send a flow of value 1 through $P$ from $v$ to $t_i$. This is not a 
maximal $v-t_i$ flow, hence there exists an augmenting path for it 
(see Bollob\'as~\cite[Chapter III]{bollob1998modern} for basic facts related to the Ford-Fulkerson
theorem). To form such a path we can use the arcs of $P$ in the reverse   
direction and all other arcs of $G^*_i$ in their original direction. In
fact,
it suffices to search for an augmenting path until the first node ($m$ in $G^*_3$ on 
Fig. \ref{fig:non-crossing-cuts}(b)) incident to $P_1$ or $P_2$ is accessed (say, it is $P_2$). From $m$ one
can walk along the arcs of $P_2$ to reach $t_i$. Let this $m-t_i$ path be 
$R$ (e.g. $e_2$ on  $G^*_3$). Suppose that from $v$ up to $m$ we used the
arcs $h_1,\ldots ,h_l\in E$ in addition to some edges which are reverses of
arcs in $P$. Then the resulting flow of value 2 from $v$ to $t_i$ gives two 
arc disjoint $v-t_i$ paths $Q \cup R$ and $Q'$ 
(e.g. $Q \cup R = \{ h_1 \} \cup \{ e_2 \}$ and $Q' = \{ e_3 \}$ in $G^*_3$). This can be shown by a routine parity argument, such as in Suurballe's algorithm. Here $Q$ is a $v-m$ path and all arcs in $Q\cup Q'$ are from the arc set $P\cup \{h_1,\ldots ,h_l\}$.  

We set the dominant arc set $E^i_1$ as $E^i_1=P_2\cup \{g,Q,R\}$, and other    
two arc sets will be $P_1$, and $\{g,Q'\}$.  

Type~(iv) is similar to Type~(iii), essentially the same
argument works with a (reverse) flow of value 3 from $t_i$ to $s_i$.

This finishes the local parts of the construction. We claim next that 
the arc sets $E^{i}_j$ and $E^{i+1}_j$  ($j=1,2,3$) can be meaningfully glued 
together to obtain $E_1,E_2$ and $E_3$.  This is done as follows: an arc of
the form $(s_i,v)$ or $(v,t_i)$ is replaced by the respective arcs 
of $G^*$ they are obtained from. There is one ambiguity here when the cut is a 2-edge-cut with edges $f$ and $g$. Then $E^i_1$ and $E^{i+1}_1$ is joined along an arc from $f$ and along an arc from $g$.  By doing this at
every $i$, finally we have 3 arc disjoint arcs sets
$E_1,E_2,E_3$ in $G^*$, which connect $s$  to $t$. This finishes the proof.
\end{IEEEproof}
\subsection{The Challenge Behind Theorem~\ref{mainthm}}
To understand the difficulties of an algebraic proof for the RFD problem, first notice that in a graph that has a minimal cut of $3$ edges, the Ford-Fulkerson theorem implies that there are three edge-disjoint paths in the network between the source and the destination. This implies that diversity coding suffices to ensure instantaneous recovery. However, our result implies instantaneous recovery even in cases where the graph has a $2$-edge-cut, whereas it still has some \emph{redundant} edges, i.e., edges whose deletion does not affect the minimal cut of the graph. We next explain why this case is non-trivial. 

Consider a case where a graph has a 2-edge-cut, whereas it has some redundant (non-minimimal cut) edges. In this case, first notice that random linear network coding suffices to ensure instantaneous recovery. 
It is instructive to understand this algebraically. In the algebraic approach of Koetter and Medard \cite{koetter2003algebraic}, the overall network transfer matrix $\mathbf{M}$ is related to the \emph{local} coding matrix $\mathbf{F}$ as $\mathbf{M}=(\mathbf{I}-\mathbf{F})^{-1}$. The local coding matrix describes the choice of coding co-efficients on every edge of the matrix. The transfer matrix between the source and destination is a sub-matrix of the overall network transfer matrix $\mathbf{M}$.  Now, the removal of an edge has been shown to be equivalent to a row reduction on the overall network transfer matrix in \cite{zeng2012edge}. If the edge does not affect the minimal cut, then it translates to a row operation on the transfer matrix between the source and the destination that does not affect its rank (when random coding co-efficients are used, and the field size is sufficient). Since the rank (which has value $2$) is preserved, instantaneous recovery is automatically ensured. 

While random linear coding ensures instantaneous recovery, doing the same when the coding solution is restricted to just splitting and merging is not simple. Algebraically speaking, restricting solutions to splitting and merging, this translates to specific constraints on the local coding matrix $\mathbf{F}$. However, when such restrictions are placed, in general, removal of an edge that does not affect the minimal cut of the graph could still affect the rank of the transfer matrix, and therefore affect instantaneous recovery. (To see this, observe that even in a routing solution, removal of an edge on a route reduces the rank of the solution by one, even if the minimal cut of the graph is not changed). 

To understand our solution from a different perspective, notice that when a redundant edge is removed, there exists an alternate decomposition of the graph into $2$ edge independent paths that could be used. A first approach that might appear to work is to combine all these various decompositions, each decomposition formed by removing one redundant edge in the network, via splitting and merging. However, such a combination is not trivial since an edge that carries $A$ in one decomposition, may be forced to carry $B$ in another, and there is no natural method to combining these solutions. The chief contribution of our paper is to ensure instantaneous recovery as long as the number of data parts is two and the number of edge failures is as at most one. 
\section{Linear Time Algorithm in Feasible Graphs}
\label{sec:noncritical}
The flow decomposition (i.e., coding) algorithm is built on the following two observations. First, \emph{finding a fixed number of arc-disjoint (or edge-disjoint) paths can  be done in $O(|E|)$ time}. It is because, the augmenting paths in the residual graph can be found with breadth-first search (BFS). Second, \emph{finding a maximal chain of minimum $s-t$ cuts $C_i$ in $G\rq{}$ can be done in linear time}, if the max flow of $G\rq{}$ has value 3. Let $G_r\rq{}$ denote the residual graph $G\rq{}$ built up by the Ford-Fulkerson algorithm, after the flow of value 3 is found. The node set $C_{i+1} \setminus C_i$ in $G_r\rq{}$ is strongly connected for $i=1,\dots,k-1$, because it has no minimal $s-t$ cuts. Identifying every strongly connected component of $G_r\rq{}$ takes $O(|E|)$ time \cite{aspvall1979linear}. Next we search for the topological ordering over the strongly connected components, which is feasible in linear time in DAGs, to obtain a maximal chain of minimum $s-t$ cuts $C_i$ in $G\rq{}$.

A closer inspection of the proof of Theorem~\ref{mainthm} reveals that we do not use in full force the fact that $G$ is critical. In fact, the following three assumptions are sufficient to carry out the construction of the arc sets $E_1,E_2,E_3$:
\begin{enumerate}[1)]
	\item $G$ is a feasible network such that $G'$ has a maximum flow of value 3.
	\item Every arc of $G'$ carries a flow value $0.5$, or $1$ or $1.5$.
	\item We have a maximal chain
$C_1 \subset C_2\subset \cdots \subset C_k $ of minimum $s-t$ cuts in $G'$ in such a way that there is no edge $(u,v)$ of capacity $1.5$ with both $u$ and $v$ belonging to $C_{j+1} \setminus C_j$ for some $j$.
\end{enumerate}

We claim that from arbitrary feasible $G=(V,E,c)$ such that the maximum flow in $G'$ is 3 we can construct a subnetwork with Properties 1)-3) in \emph{linear time}. Thus, an initial and expensive (quadratic time, to be specific) computation of a critical subnetwork can be dispensed with.

First, we compute a 3 value $s-t$ flow in $G'$. If the maximal flow is more than 3, then DC is a feasible solution, as three disjoint paths exist in the network. Thus, in the rest of this subsection we assume that the maximal flow is exactly 3. Next we find a maximal chain $\{C_j\}$ of cuts in time $O(|V|+|E|)$ as indicated previously. We delete edges with flow 0, and scan the edges $e=(u,v) $ of $G'$ which violate Property 3), and decrease the reduced capacity of $e$ to 1.

We have to show that the resulting network $G''$ (with the modified reduced
capacities) still allows an $s-t$ flow of value 3. Indeed otherwise $G$ had a cut consisting of two edges, one of them (say $e$) with capacity lowered to 1. These two edges must form a cut of value 3 in $G'$. This is impossible, because in the residual graph for the 3 value flow of $G'$ there is a directed path from $u$ to $v$, hence $e$ can not be a part of a minimum cut there.

We have to compute a 3 value flow in $G''$ and possibly a new chain of minimal cuts $C_j$. We discard the arcs with 0 flow. Note that in $G''$ any edges of capacity $1.5$ are in a minimal cut, because all the edges with $1.5$ capacity are belong to a 2-edge-cut\footnote{It is easy to see that in $G'$ there are no edges $e=(u,v)$ of capacity $1.5$ for which $u \in C_i, v \in C_j$, and $j < i$.}. The network $G''$ and the new chain of cuts satisfies Properties 1)-3). Thus, using $G''$ one can construct arc sets $E_1,E_2,E_3$ by following the theoretical construction of Theorem~\ref{mainthm} along the chain of cuts $\{C_j\}$ in time $O(|V|)$.
\section{Conclusions}
\label{sec:conclusions}
In this paper we proved that the single edge failure resilient protection for connections where data flow can be split into two subflows can be decomposed into three end-to-end resilient DAGs. The importance of our results is twofold. First, we gave a new structure theorem (Theorem~\ref{mainthm}) for feasible networks, while we employ no network transformations. Second, we showed that a resource efficient protection method with instantaneous recovery in $2$-connected topologies with some redundant edges -- where the yet efficient diversity coding approach fails-- is feasible without the modification of the core switches in transport networks. 
\section*{Acknowledgments}
P. Babarczi was supported by the J\'anos Bolyai Research Scholarship of the Hungarian Academy of Sciences (MTA). Research of P. Babarczi and J. Tapolcai was partially supported by the Hungarian Scientific Research Fund (OTKA grant K108947). L. R\'onyai was supported by the Hungarian Research Fund (OTKA grants NK105645, K77476) and T\'AMOP-4.2.2/b-10/1-2010-0009. This document has been produced with the financial assistance of the European Union under the FP7 G\'EANT project grant agreement number 605243 as part of the MINERVA Open Call project.
%
%
%

\end{document}